\documentclass[journal=jacsat,manuscript=article]{achemso}

\usepackage{chemformula} 
\usepackage[T1]{fontenc} 
\usepackage{tabularx}

\author{B. Chettri}
\affiliation[Pachhunga University College]
{Physical Sciences Research Center (PSRC), Pachhunga University College, Department of Physics, Aizawl, 796001, Mizoram, India}
\alsoaffiliation[North-Eastern Hill University]
{Department of Physics, North-Eastern Hill University, Shillong, Meghalaya, 793022, India}
\author{P. K. Patra}
\affiliation[North-Eastern Hill University]
{Department of Physics, North-Eastern Hill University, Shillong, Meghalaya, 793022, India}
\author{D. P. Rai}
\affiliation[Pachhunga University College]
{Physical Sciences Research Center (PSRC), Pachhunga University College, Department of Physics, Aizawl, 796001, Mizoram, India}
\email{dibya@pucollege.edu.in}

\title[An \textsf{achemso} demo]
  {Enhanced \ch{H$_2$} storage capacity of the bilayer hexagonal Boron Nitride(h-BN) incorporating Van der Waals interaction under applied external electric field}

\keywords{DFT, Hydrogen storage, adsorption energy, Gravimetric density \LaTeX}

\begin{document}







\begin{abstract}
 Light weight 2D materials due to its large surface area are being studied for its hydrogen storage applications. The characteristics of hydrogen adsorption on the electric field induced h-BN bilayer were investigated. The overall storage capacity of the bilayer is 6.7 wt\% from our theoretical calculation with $E_{ads}$ of 0.308 eV/H$_2$. The desorption temperature for H$_2$ molecules from the h-BN bilayer system in the absence of external electric field is $\sim$ 243 K. With the introduction of external electric field, the $E_{ads}$ lies in the range 0.311-0.918 eV/H$_2$ with a desorption temperature of 245-725 K. The charge transfer analysis reveals that 0.001-0.008 |e| of electronic charge transfer between the h-BN bilayer and H$_2$ molecules. Henceforth, the adsorption mechanism is through a weak Van der waals interaction. Our results shows that the external electric field enhances the average adsorption energy as well as the desorption temperature and thus making the h-BN bilayer a promising candidate for hydrogen storage.
\end{abstract}

\section{Introduction}
Fossil fuels which dominates the entire world for energy sources has major drawbacks due to its high contribution on the environmental pollution and global warming due to the emission of CO$_2$ and other harmful pollutants. Hydrogen can be an alternative to the fossil fuels as it is abundant in nature, has high energy density and doesn't emit any harmful pollutants under combustion, is a high efficiency energy carrier. The quest for light weight solid state materials for hydrogen storage is at peak as the conventional storage techniques has some major drawbacks. For a material to be an efficient hydrogen storage media, there are certain benchmark criteria set up by the United States Department of Energy i.e., gravimetric density should be greater than 6.0 wt\% and average adsorption energy $\sim$ 0.2-0.8 eV per H{$_2$}.
2D materials like graphene, silicene, borophene, MoS$_2$ etc., are light weight which is advantageous for high gravimetric density and provides large surface area which is suitable for more number of hydrogen molecules adsorption \cite{Naqvi2018,Hussain2015,Song2014,Zheng2020,Wang2016,Rai2020,Rai2020b}. The above studies also suggested that, hydrogen molecules has low binding affinity on pure systems and thus reducing the possibility to be a promising hydrogen storage media. Hussain et al., employed the density functional theory and reported that graphane under alkali and alkaline metal decoration behaves as a promising hydrogen storage candidate \cite{Hussain2011,Hussain2012a}.\\
Gao et al. successfully synthesized the mono-layer, bilayer and multi-layer h-BN on Pt foils \cite{Gao2013}. Later, Uchida et al. were also able to successfully grow large area multi-layer h-BN for further utilisation in gate-insulating material and other practical purposes \cite{Uchida2018}. h-BN bilayer has been studied for its potential application in thermoelectric devices due to its superior thermal conductivity \cite{Wang2016b} and for nano-electronic, opto-electronic devices by modulating the energy band gap by applying external electric field, strain, doping etc., \cite{Abergel2015,Amiri2020,Chettri-2021c,Laturia2018,Ansari2015}. Extensive research on h-BN mono-layer, graphitic carbon nitride, boron carbide mono-layer's potentiality reversible hydrogen storage system for mobile applications are in progress \cite{Hussain2016,Alhameedi2019,Hu2014,Hu2021,Li2013,Li2017,Mananghaya2019,Naqvi2017}. Panigrahi et al., reported co-adsorption innovative technique for hydrogen storage of lithiated carbon nitride (C{$_7$}N{$_6$}). They reported that, co-mixing of CH{$_4$}-H$_2$ on carbon nitride(C{$_7$}N{$_6$}) enhances the gravimetric density to 8.1 wt\% which is beyond the criteria set by USDOE \cite{Panigrahi2021}. Chettri \textit{et. al.}, reported the most probable H$_2$ adsorption site to be hollow site and maximum H$_2$ uptake capacity of the pristine h-BN mono-layer to be 6.7 wt\% with an average adsorption energy of 0.13 eV/H$_2$ \cite{Chettri2021}. Hu et al. performed DFT based comparative study for hydrogen storage properties of lithium decorated h-BN/graphene hybrid domains and reported the gravimetric density upto 8.7 wt\% \cite{Hu2014}. Recently, Banerjee et al., studied about the 5.5-11.1\% lithium fictionalised on hydrogenated hexagonal boron nitride mono-layer using DFT implemented in VASP. Lithium fictionalization enhances the gravimetric density($\sim$ 6 wt\%) as well as the average adsorption energy favorable for adsorption/desorption process \cite{Banerjee2016}. Hussain et al. highlighted the role of strain on the lithium fictionalized grapahane to enhance the hydrogen storage capacity up to 12.12 wt\% \cite{Hussain2012e}. Application of external electric field  enhances the hydrogen storage ability on Calcium decorated-silicene mono-layer and bilayer \cite{Song2014}. As reported earlier, external electric field tunes the CO$_2$ molecules adsorption on h-BN nanosheet \cite{Sun2013}. Also, ZnS mono-layer and bilayer band gap tuning has been successfully reported \cite{Rai2018}.
As per our knowledge, the influence of applied external electric field on hydrogen storage properties of pristine h-BN bilayer and h-BN bilayer has not been reported till date. Motivated by the above work, we present our DFT study of h-BN bilayer for hydrogen storage applications in a field-free condition and in the presence of external electric field.

\section{Computational detail}
All the calculations were performed using density functional theory implemented in SIESTA code \cite{Kohn1965, Soler_2002}. We used generalized gradient approximations for the exchange-correlation functional with Perdew-Burke-Ernzerhof(PBE) \cite{Perdew1996}. We generated a norm conserved Troullier-Martins van der Waals pseudo-potentials for B, N and H atoms in order to describe the interactions between core and valence electrons using ATOM program  \cite{Troullier1991,Soler_2002}. The pseudo-potentials are well tested before employing it in our calculations. Since, GGA(LDA) underestimates(overestimates) the electron interaction energies. Hence, to accurately describe the electron interaction energies PBE-GGA with vdW-LMKLL(equivalent to DFT-D2) is employed for our calculations \cite{Lee2010,Hussain2014,Hussain2017,Panigrahi2020b,Buimaga-Iarinca2014}.  A double zeta  polarization basis set was used to expand the wave-functions of the valence electrons. A mesh cut off of 450.0 Ry was used for k-grid integration. A 3$\times$3 super-cell of BN nanosheet is constructed with 18(9Boron, 9 Nitrogen) atoms in each layer maintaining Boron-Nitrogen ratio of 1:1. The distance between 2 bilayer system is kept at 25 {\AA} to avoid the interaction between them. 
The conditions maintained for geometry optimization using the conjugate gradient(CG) algorithm are i) maximum force component less than 0.02 eV/{\AA} on individual atoms and ii) the total energy changes were below $10^{-5}$ eV. For all electronic calculations, we have used K-mesh of 24$\times$24$\times$1 within Monkhorst package \cite{Monkhorst1976}. 
We supposed two perpendicular directions normal to the h-BN bilayer system for an applications of external electric field i) along the +Z axis (positive field), ii) along the -Z axis (negative field) in the range $\pm$ 2 V/{\AA} with an increment of 0.2 V/{\AA}.\\
To understand the adsorption of H$_2$ molecules on h-BN bilayer, the quantitative value of binding energy ($E_{BE}$) is obtained using \cite{Varunaa2019}
\begin{equation}
E_{BE}={E_{BN}+n{E_{H2}}-{E_{nH2+BN}}}
\label{eq1}
\end{equation}
An average adsorption energy ($E_{ads}$ eV/H$_2$) of adsorbed H$_2$ molecules on h-BN bilayer system \cite{Varunaa2019}, 
\begin{equation}    
E_{ads}(eV/H2) = \frac{{E_{BN}}+n{E_{H2}}-{E_{nH2+BN}}}{n}
\label{eq2}
\end{equation}
$E_{BN}$,$E_{H2}$ and $E_{nH2+BN}$ are the total energies of pristine h-BN bilayer, an isolated H$_2$ molecule, H$_2$ molecules adsorbed h-BN bilayer. The average adsorption energy is calculated in external field-free and in presence of external field condition.\\ 
The hydrogen storage capacity is calculated in terms of gravimetric density using the following equation \cite{Chettri2021,Varunaa2019},
\begin{equation}
H_{2}(wt\%)=\frac{nM_{H2}}{(nM_{H2}+M_{BN+Li})}\times100
\label{eq3}
\end{equation}
where, $M_{H2}$ $\rightarrow$ atomic mass of H$_2$, $M_{BN}$ $\rightarrow$ atomic mass of h-BN bilayer and $n$ denotes the number of adsorbed H$_2$ molecules.\\
Similarly, the desorption temperature (T$_{D}(K)$) of the H$_2$ molecules is computed using the following equation \cite{Chettri2021},
\begin{equation}
T_{D}(K)={\frac{E_{ads}}{K_{B}}}{\left({\frac{\Delta S}{R}} - ln P \right)^{-1}}
\label{eq4}
\end{equation}
where, ${E_{ads}}$, $R$ = 8.31 ${J{{K^{-1}}{Mol^{-1}}}}$, ${K_{B}}$ = ${1.38\times10^{-23}}{J{K^{-1}}}$, $\Delta${$S$} = 75.44 ${J{{K^{-1}}{Mol^{-1}}}}$ and $P$ indicates an average adsorption energy, gas constant, Boltzmann Constant and change in the H$_2$ entropy from gas to liquid phase at the equilibrium pressure P=1 atm, respectively.\\	
The recovery rate is calculated using following equation
\begin{equation}
\tau = {{\nu_{0}^{-1}}{e}^{-{E_{ads}}/{K_{B}}T}}
\label{eq5}
\end{equation}
where ${\nu_{0}}$ is the attempted frequency($10^{12} s^{-1}$), $E_{ads}$ is the adsorption energy. We have calculated the recovery rate considering temperature(T) = 300 K.
\section{Result and discussion}
\begin{figure*}[!htb]
   	\includegraphics[width=13.00cm,  height=14.00cm]{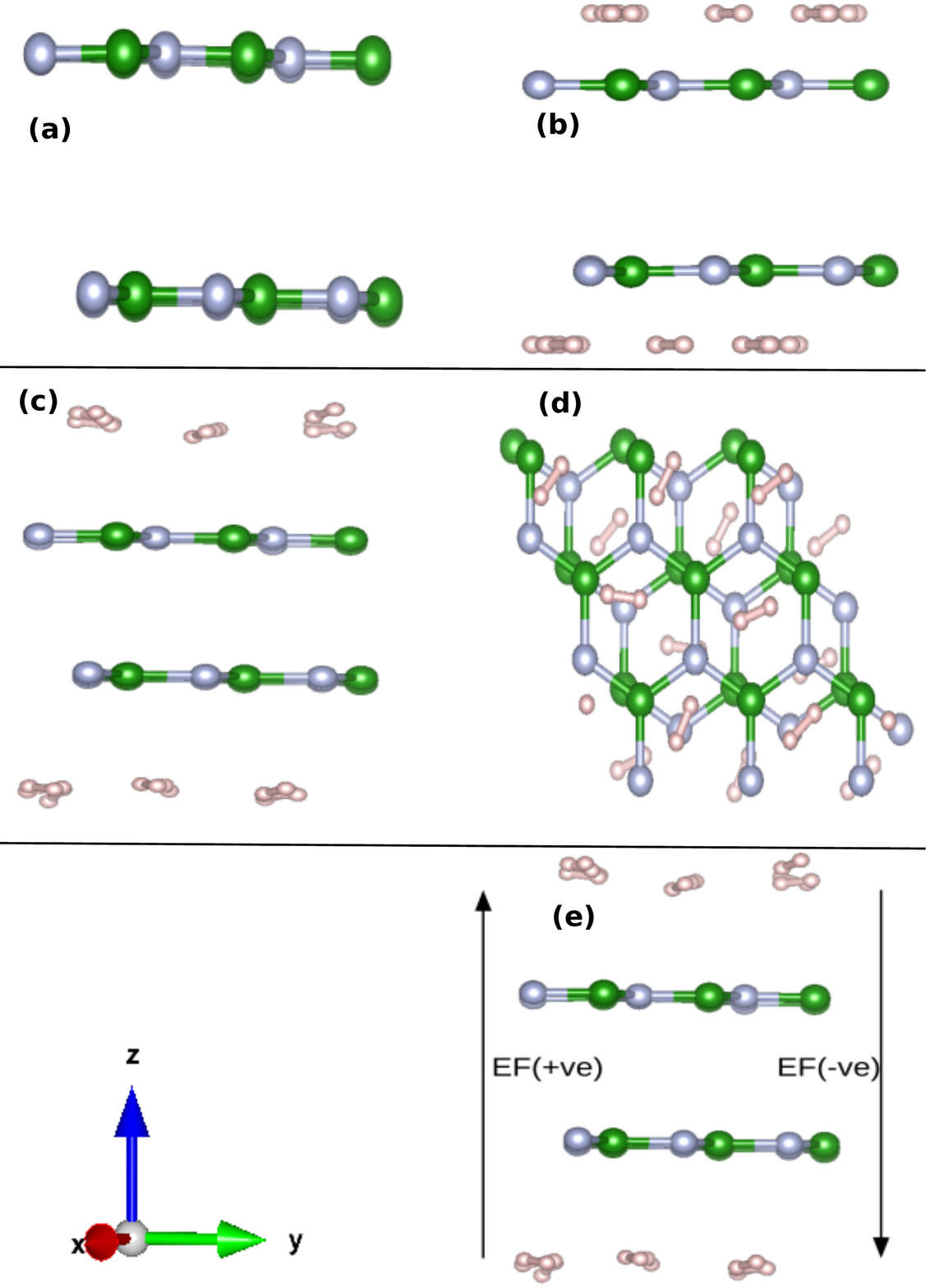}
   	\caption{\label{fig:str-final} Optimized atomic structure of h-BN bilayer is presented in Fig.(a). Similarly, Fig.(b) is the hydrogen adsorbed bilayer, (c)-side and (d)-top view of the optimized hydrogen adsorbed h-BN bilayer system. Final figure (e) shows direction of the applied positive[$E_F$(+ve)] and negative[$E_F$(-ve)] external electric field($E_F$) directions. The green,grey and brown spheres are Boron, Nitrogen and Hydrogen atoms, respectively.}
   \end{figure*}

In this section, the structural parameters and electronic properties of the pristine hexagonal boron nitride bilayer stacked in AB configuration has been discussed. The optimized atomic structure of the h-BN bilayer is presented in Fig.\ref{fig:str-final}(a). The optimized inter-layer distance is 3.34 {\AA}. The bond length between B-N is found to be 1.45 {\AA}. The band structure calculation depicts the large band gap semiconductor nature of h-BN bilayer with a direct energy band gap of 4.23 eV. These results agrees well with the earlier reported results \cite{Amiri2020,Chettri2021a,Chettri-2021c}. Observation of band structure diagram and DOS plot revealed that the B-2p states populated the conduction band and N-2p states populated the valence band, respectively. Electron density distribution plot in Fig.\ref{fig:ED} indicates the ionic-covalent type bonding characteristics between B(green)-N(grey) in a h-BN bilayer system as the charges are mostly localised on N-atom and feeble distribution between the B-N bonds.
	\subsection{Adsorption of H$_2$ on pristine h-BN bilayer in an external field-free condition}
\begin{figure*}[!htb]
			\includegraphics[width=16.50cm,  height=10.00cm]{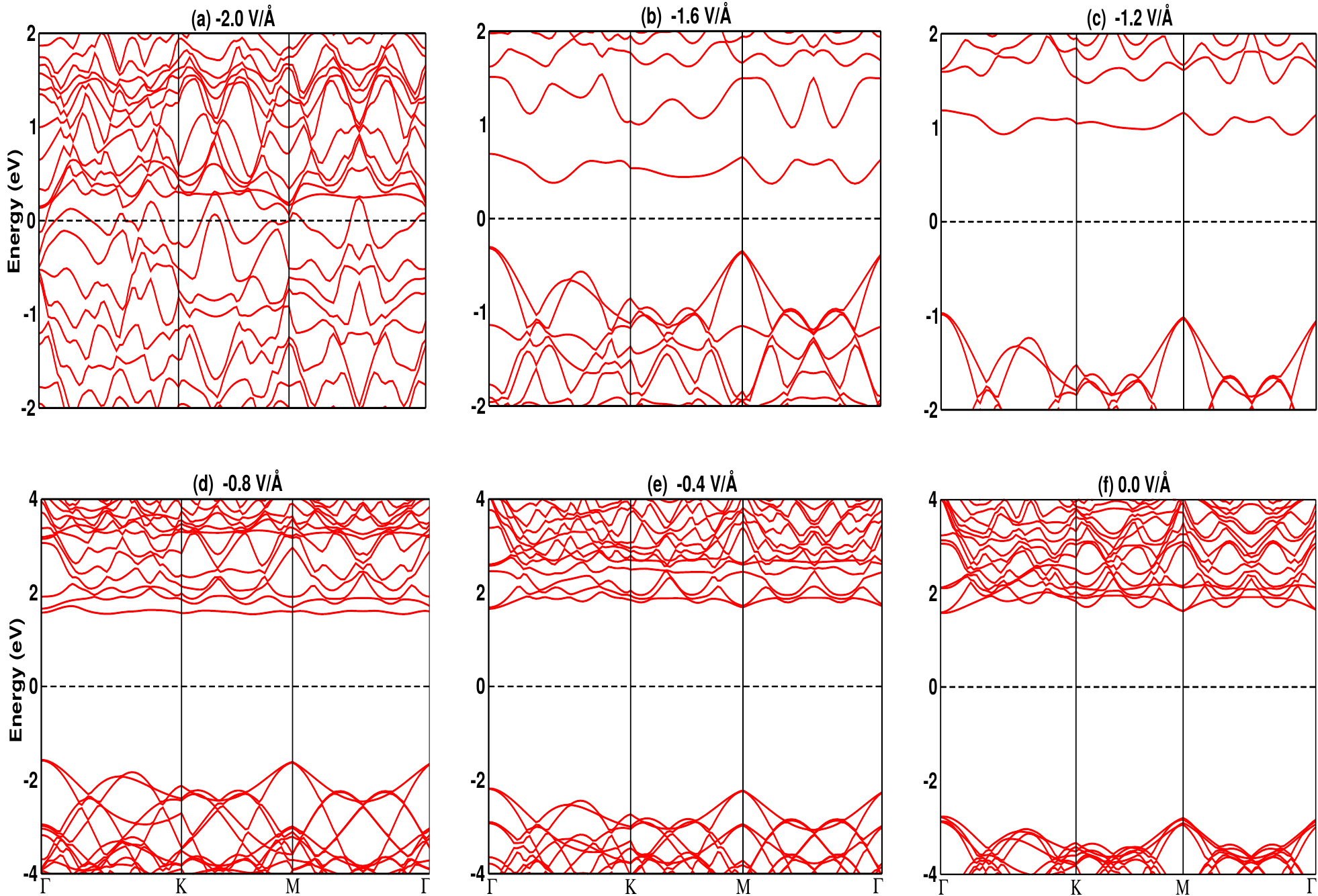}
			\caption{\label{fig:bands-neg}Band structures of the hydrogen adsorbed h-BN bilayer with electric field -2.0 to 0.0 with a difference of 0.4 and specifically labelled in each plot. In this case, direction of the electric field is from top layer to bottom layer. The Fermi level is denoted by dotted horizontal line at 0 eV.}
		\end{figure*}

\begin{figure*}[!htb]
			\includegraphics[width=16.50cm, height=10.00cm]{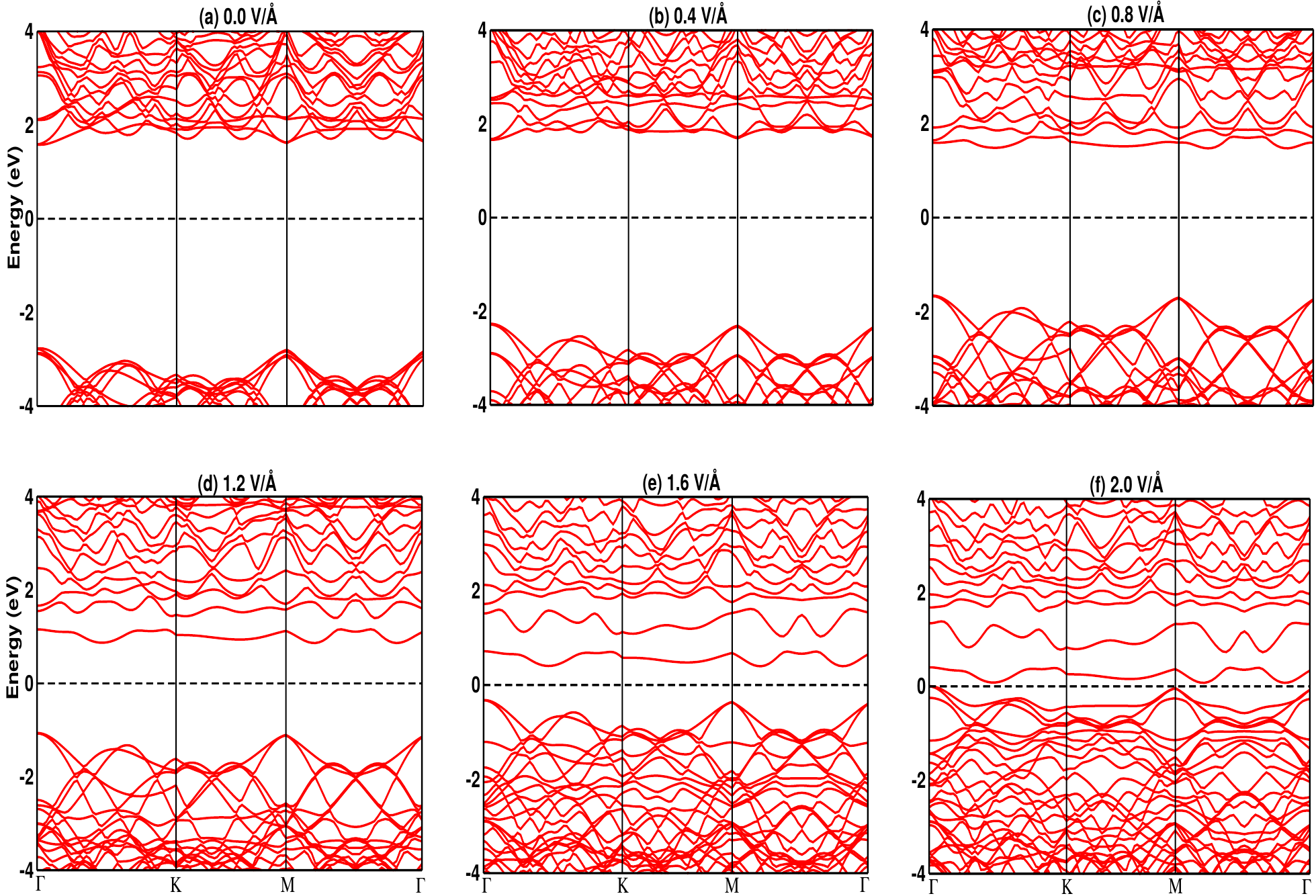}
			\caption{\label{fig:bands-pos}Band structures of the hydrogen adsorbed h-BN bilayer with electric field 0.0 to 2.0 with a difference of 0.4 and specifically labelled in each plot. In this case, direction of the electric field is from bottom layer to top of the h-BN bilayer. The Fermi level is denoted by dotted horizontal line at 0 eV.}
		\end{figure*}

	This section is dedicated for the discussion about the adsorption of H$_2$ molecules on the pristine hexagonal boron nitride bilayer system. As observed from the optimized atomic structure, H$_2$ molecules tends to reside on the hollow hexagonal site of the h-BN bilayer. Similar observation about the hollow hexagonal site of being the probable site for hydrogen adsorption has been reported for graphene, h-BN nanosheet etc., \cite{Chettri2021}. We haven't followed a fabrication process to build the reservoir of H$_2$ molecules into the h-BN bilayer as there in no risk of clusterization of H$_2$ molecules at the surface or migration on the surface of host materials in the absence of decoration or doping with other atoms. Initially, all H$_2$ molecules were oriented parallel to the plane of h-BN bilayer and under optimisation some of the H$_2$ molecules tilted. With the introduction of hydrogen molecules, inter-layer spacing increased to 3.67 {\AA}. The optimised average vertical distance of the H$_2$ molecules on the top and bottom layer is in between 2.78-3.1 {\AA}. Free H$_2$ molecule has a bond length of 0.74 {\AA}. Under adsorption, H-H bond length varies between 0.74-0.75 {\AA} and remains in molecular form \cite{Lopez-Corral2011}. The effect of hydrogen molecule adsorption on the electronic band structure of h-BN bilayer has been presented in Fig.\ref{fig:bands-neg}(f), \ref{fig:bands-pos}(a). The density of states(DOS) of hydrogen molecules adsorbed h-BN bilayer in terms of B-2p, N-2p, H-1s orbital contribution denoted by black coloured states(0.0) is presented in Fig.\ref{fig:dos-neg}, \ref{fig:dos-pos} (a-c). As evident from the DOS plot, the valence and conduction bands is dominated by the N-2p and B-2p orbital respectively [see Fig.\ref{fig:dos-neg},\ref{fig:dos-pos}(a-b)]. Whereas, H-1s orbital contribution is observed beyond 2 eV energy range [see Fig.\ref{fig:dos-neg},\ref{fig:dos-pos}(c)].
	
\begin{figure*}[!htb]
				\includegraphics[width=16.50cm, height=6.00cm]{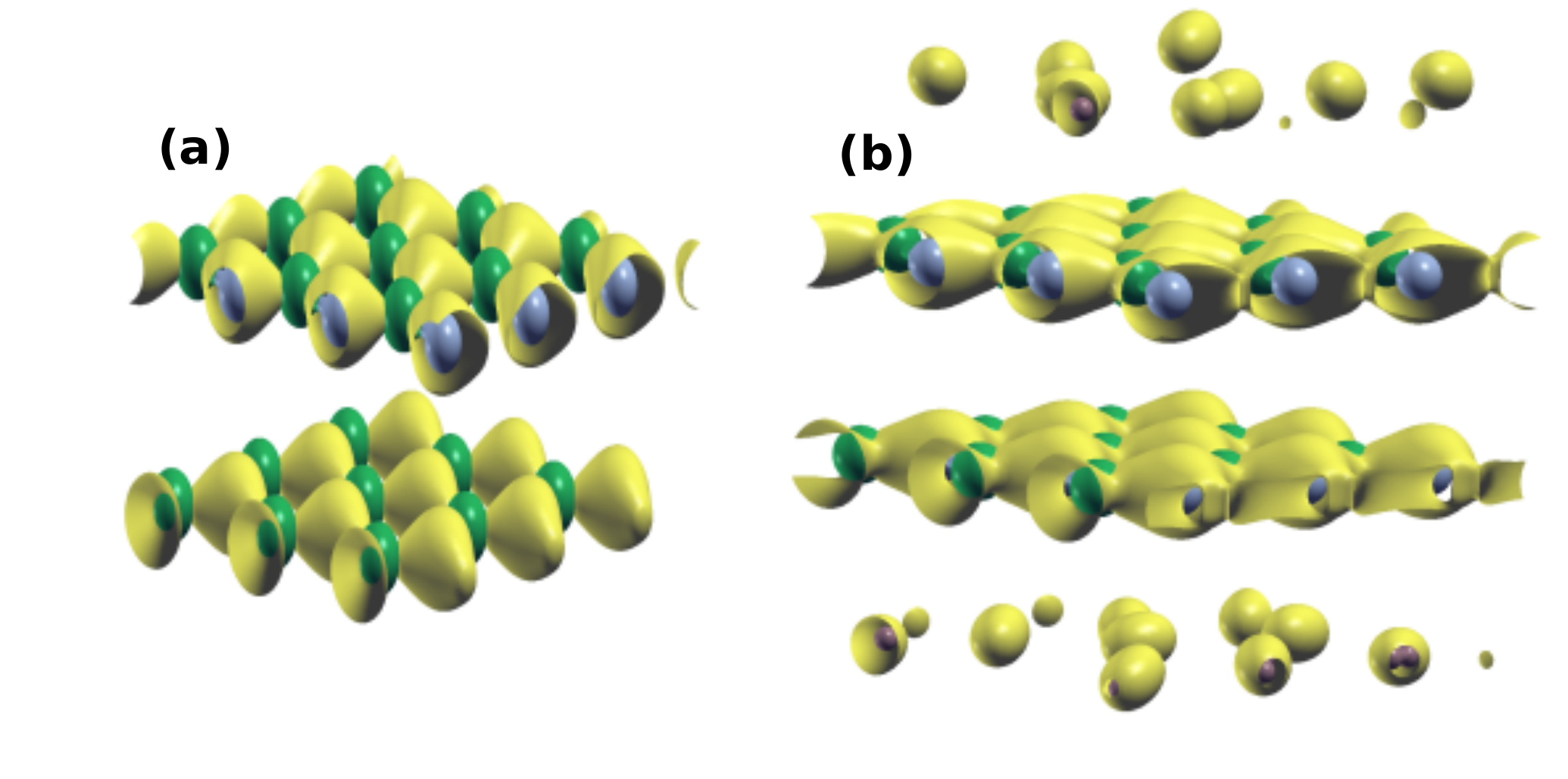}
				\caption{\label{fig:ED} Electron density plot of the optimized (a) pristine h-BN bilayer and (b) 16$H_2$ adsorbed h-BN bilayer. The green, grey and brown spheres are boron, nitrogen and hydrogen atoms, respectively. Yellow lobe shows the electron density distribution.}
			\end{figure*}

	The energy band gap in an external field-free condition is 4.31 eV [see Fig.\ref{fig:bands-neg}(f),\ref{fig:bands-pos}(a)]. 
	The electron density distribution of the pristine h-BN bilayer and 16H$_2$ molecules adsorbed system is presented in Fig.\ref{fig:ED}(a,b). As observed from the electron density plot, most of the electrons are localised around nitrogen atom(grey). Also, the charge transfer between the h-BN layer is absent. Furthermore, there is no overlapping of electron distribution between the h-BN bilayer and H$_2$ molecules thus validating the adsorption mechanism of H$_2$ molecules as weak van der Waals interactions. Total energies of h-BN bilayer and H$_2$ molecules adsorbed bilayer are calculated in a field-free condition. In an external field-free condition, the calculated weight percentage of H$_2$ molecules and average adsorption energy of H$_2$ molecules on h-BN bilayer are 6.7 wt\% and 0.308 eV/H$_2$ respectively. The calculated desorption temperature in this case is $\sim$243 K. The hydrogen storage capacity and average adsorption energy lies within the benchmark set by USDOE \cite{Schlapbach2010,Varunaa2019}, whereas the desorption temperature is slightly less than the room temperature. The desorption temperature can be further increased by the introduction of external electric field($E_F$).
	
 	\subsection{Effect of external Electric Field on H$_2$ adsorption on h-BN bilayer}
\begin{figure*}[!htb]
				\includegraphics[width=16.50cm, height=6.00cm]{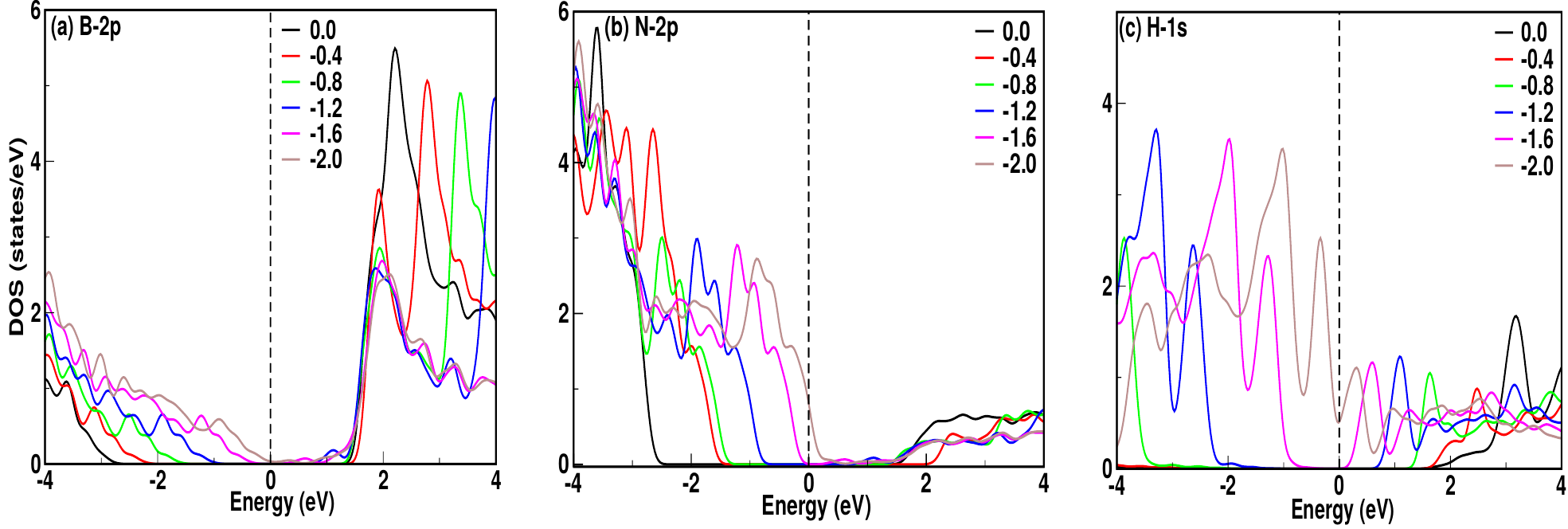}
				\caption{\label{fig:dos-neg}Density of States(DOS) plot of the B-2p, N-2p, H-1s orbitals with electric field -2.0 to 0.0 with a difference of 0.4 and specifically labelled in each plot. In this case, direction of the electric field is from top layer to bottom layer. The Fermi level is denoted by dotted vertical line at 0 eV. }
			\end{figure*}

\begin{figure*}[!htb]
				\includegraphics[width=16.50cm, height=6.00cm]{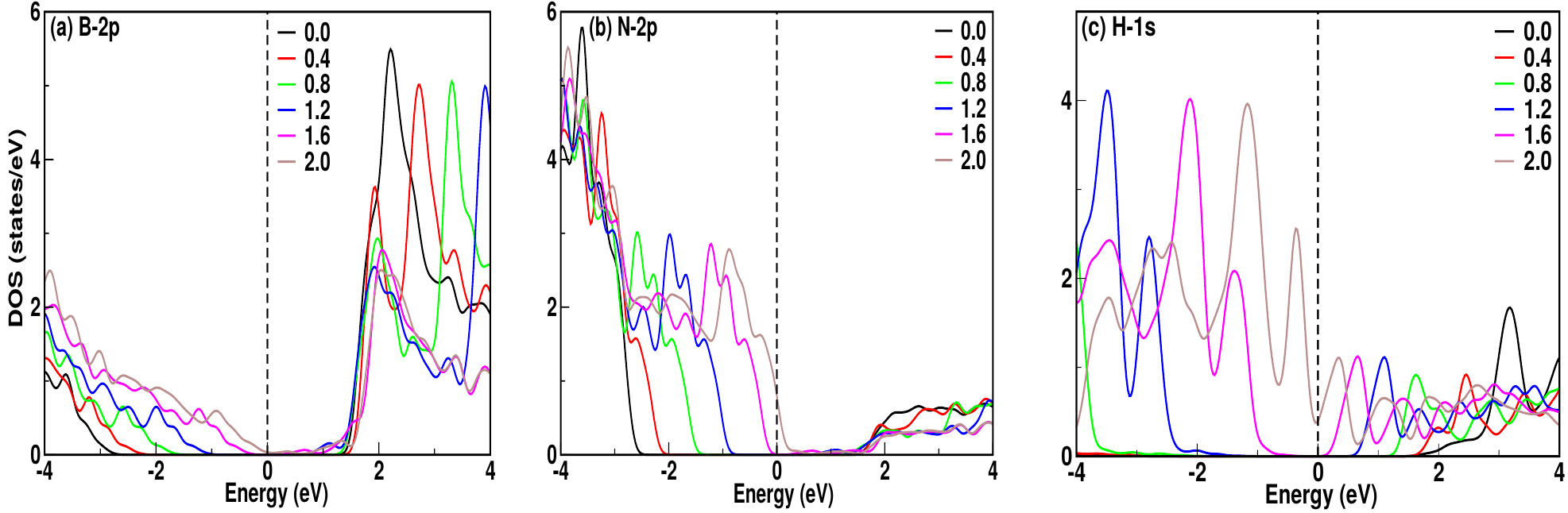}
				\caption{\label{fig:dos-pos}Density of States(DOS) plot of the B-2p, N-2p, H-1s orbitals with electric field 0.0 to 2.0 with a difference of 0.4 and specifically labelled in each plot. In this case, direction of the electric field is from bottom layer to top of the h-BN bilayer. The Fermi level is denoted by dotted vertical line at 0 eV.}
			\end{figure*}

	In this section, we will discuss about the hydrogen storage ability of h-BN bilayer system in the presence of external electric field. External electric field is introduced normal to the hydrogen molecules adsorbed h-BN bilayer in the range [-2.0, 2.0 V/{\AA}] with an increment of 0.2 V/{\AA} in two direction i.e, bottom to top layer(positive) and top to bottom layer(negative) of the h-BN bilayer [see Fig.\ref{fig:str-final}(e)]. We first investigated the effect of different external field intensity on the bond-lengths of adsorbed H$_2$ molecules, inter-layer h-BN bilayer distance and the adsorbed H$_2$ molecules distance from the h-BN top and bottom layer plane. As all the adsorbed H$_2$ molecules were nearly parallel to the h-BN plane, a polarisation effect on H$_2$ was least feasible and hence the change in bond-length was minimal i.e., 0.74-0.76 {\AA} compared to free H$_2$ molecule bond-length of 0.74 {\AA}. The h-BN bilayer distance was reduced by $\sim$ 0.13 {\AA} when the applied field was $\pm$ 2 V/{\AA}.
	Band structure and PDOS has been presented in Fig.\ref{fig:bands-neg}(a-e),\ref{fig:bands-pos}(b-f) and Fig.\ref{fig:dos-neg}, \ref{fig:dos-pos}(a-c), respectively to further study the electronic properties of H$_2$ molecules adsorbed h-BN bilayer in the presence of external electric field. The quantitative analysis of variation of the energy band gap vs positive and negative external field is summarised in Table \ref{tab:Energy}. A substantial change in the band gag is observed as seen qualitatively from the band-structure and DOS plots in Fig.\ref{fig:bands-neg}, \ref{fig:bands-pos}, \ref{fig:dos-neg}, \ref{fig:dos-pos}. The strength of electric field are labelled by respective colors in the B-2p,N-2p and H-1s orbital DOS plots [see Fig.\ref{fig:dos-neg},\ref{fig:dos-pos}(a-c)]. 
	
	The effect of different electric field intensity on B-2p orbital on negative direction is presented in Fig.\ref{fig:dos-neg}[a]. On careful inspection, negative external electric field has less effect on the conduction band region whereas, on the valence band region, B-2p orbital contribution increases near the Fermi level with increase in intensity of the field in negative direction [see Fig.\ref{fig:bands-neg}(a)]. Similarly, DOS plot of N-2p orbital contribution is presented in Fig.\ref{fig:dos-neg}(b). Here, we can observe that N-2p orbital contribution is maximum in the valence band and with the increase in intensity of the external field, the contribution is shifted towards the Fermi level. Whereas, its orbital contribution in the conduction band remains minimal. Similar effect of external electric field on the H-1s orbital is observed from Fig.\ref{fig:dos-neg}(c). In all the above cases, a finite band gap is observed when the external field strength is between 0 - (-1.6) V/{\AA} and the system shows a metallic behaviour at -2.0 V/{\AA}. 	A similar conclusion can be drawn about B,N-2p and H-1s orbital contribution on the electronic properties of H$_2$ molecules adsorbed h-BN bilayer system under the application of positive external field [see Fig.\ref{fig:dos-pos}(a-c)]. The variation in the energy band gap of H$_2$ adsorbed h-BN bilayer system under applied external electric field at same intensity but in opposite direction is less than 0.13 eV.
	
   We have calculated the desorption temperature(T$_D$) for H$_2$ molecules presented in Table \ref{tab:Energy}. From eqn.\ref{eq4}, it is evident that desorption temperature is dependent on the average adsorption energy. In our study, external electric field enhances the average adsorption energy and thus increases the desorption temperature(T$_D$) [see Table\ref{eq4}]. The desorption temperature under different external electric field lies in the range 245-725 K with an average adsorption energy in the range 0.311-0.918 eV/H$_2$. Therefore, the external electric field can effectively stabilize the h-BN bilayer structure with average adsorption energy of 0.311-0.918 eV/H$_2$ and desorption at temperature 245-725 K which is near ambient temperatures. We also performed the Hirsfield charge transfer analysis to understand the H$2$ adsorption mechanism on the h-BN bilayer system in case of electric field-free and under the application of external electric field at different intensity. Charge transfer analysis using Voronoi and Hirshfeld, revealed that there is a transfer of charge of the order of 0.001-0.008 |e| between the h-BN bilayer and adsorbed H$_2$ molecules. 
   The average adsorption energy has substantially increased upon the introduction of electric field. The percent change in average adsorption energy({${\Delta}_{ads}$}) with respect to the applied external electric field has been present in Table \ref{tab:Energy}. $\pm$ 1.6 V/{\AA} has highest percent change w.r.t the external electric field. Also, the electronic property of the host material has been tuned by the application of external electric field. With the increase in external field strength, bands are shifted near the Fermi level and thus reduces the energy band gap. Under the application of external electric field, a charge redistribution between B and N atoms in respective h-BN layers is observed. The Nitrogen atoms became more negative and Boron atoms more positive. Whereas, a minimal charge redistribution are observed in H-atoms. Thus, the charge redistribution may have created an electrostatic attraction between the bilayer and reduces the h-BN bilayer distance upon external electric field. Also, it may create an attractive force between the H$_2$ molecules and h-BN bilayer, thus tuning the adsorption energy.
   The external positive and negative electric field have similar effect on the energy band gap(E$_g$) and the average adsorption energy(E$_{ads}$)[see Table\ref{tab:Energy}].
	 	
\begin{table}[htb!]
	\caption{The calculated Energy Bandgap (E$_{g}$), Average adsorption energies (E$_{ad}$), the percent change of the average adsorption energy ({$\Delta _{ads}$}\%) induced by external electric field, Desorption Temperature (T$_D$) and recovery rate($\tau$) of the hydrogen adsorbed hexagonal boron nitride bilayer system under the influence of external electric field($E_F$).}
	\label{tab:Energy}	
	{%
			\begin{tabular}{cccccccccccccccc}
				\hline
				$E_F$ (V/\AA) &  {E$_{g}$} (eV) & {E$_{ads}$} (eV/H$_2$) & {$\Delta _{ads}$}\%	& wt\% & {T$_{D}$} (K) & Recovery rate ($\tau$) (s) &	 \\
				\hline
				-1.6      &  0.67 &   0.918  & 198.051 & 6.7 &  725.960 & 2941 s &\\
				-1.2      &  1.98 &   0.369  & 19.805  & 6.7 &  291.954 & 1.66  $\mu$s &\\
				-0.8      &  3.13 &   0.333  &  8.116  & 6.7 &  263.218 & 406 $\eta$s &\\
				-0.4      &  3.82 &   0.315  &  2.272  & 6.7 &  249.195 & 204  $\eta$s &\\
				 0.0   	  &  4.31 &   0.308  &   ---   & 6.7 &  243.468 & 154  $\eta$s &\\
				 0.4      &  3.90 &   0.311  & 0.974   & 6.7 &  245.887 & 193  $\eta$s &\\
				 0.8  	  &  3.24 &   0.329  & 6.818   & 6.7 &  260.206 & 350  $\eta$s &\\
				 1.2 	  &  1.93 &   0.355  & 15.259  & 6.7 &  280.450 & 945  $\eta$s &\\
				 1.6  	  &  0.70 &   0.666  & 116.233 & 6.7 &  526.386 & 0.164 s &\\
				\hline
			\end{tabular}}
		\end{table}
		\begin{figure*}[!htb]
			\includegraphics[width=16.50cm,  height=7.00cm]{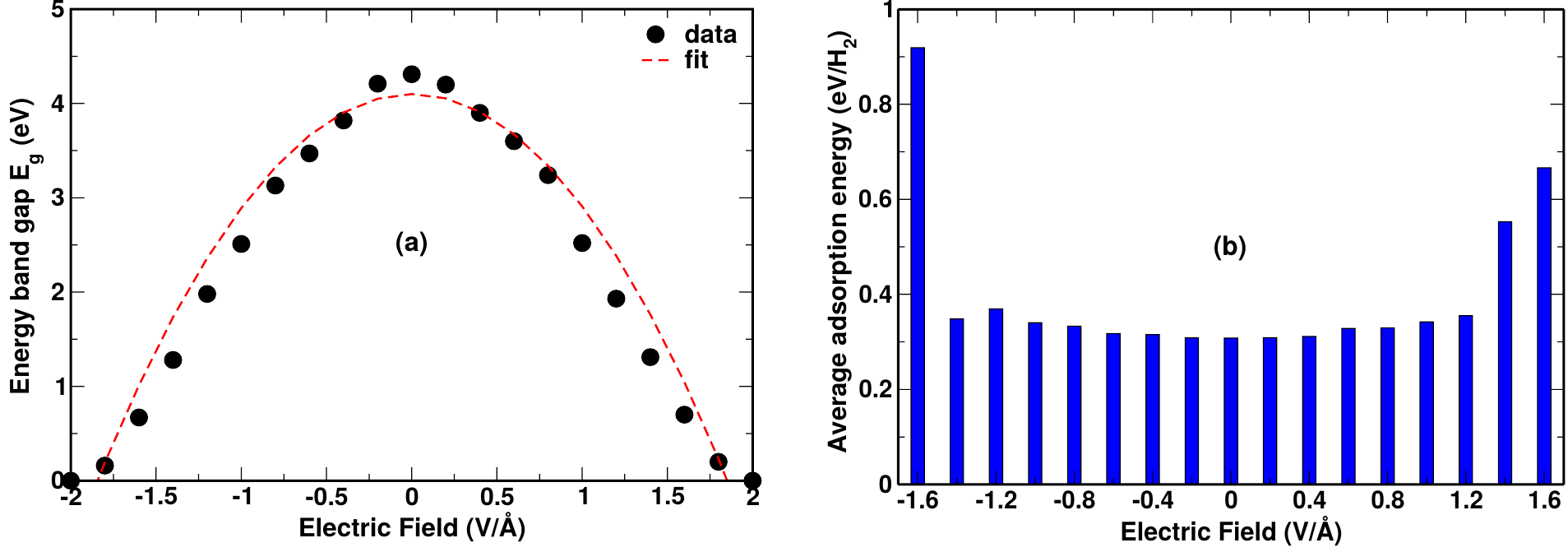}
			\caption{\label{fig:ads}(a) The variation of energy band gap with external electric field. Solid black circles represents the energy band gap with respect to external electric field. Whereas, black dotted curve is a fit with equation $y$ = 4.1013 + 0.01 $x$ - 1.20 $x^2$, here $x$ is the applied varying electric field and $y$ is the variation of bandgap. (b) The variation of average adsorption energy with respect to external electric field.}
		\end{figure*}
				\begin{figure*}[!htb]
					\includegraphics[width=16.50cm,  height=7.00cm]{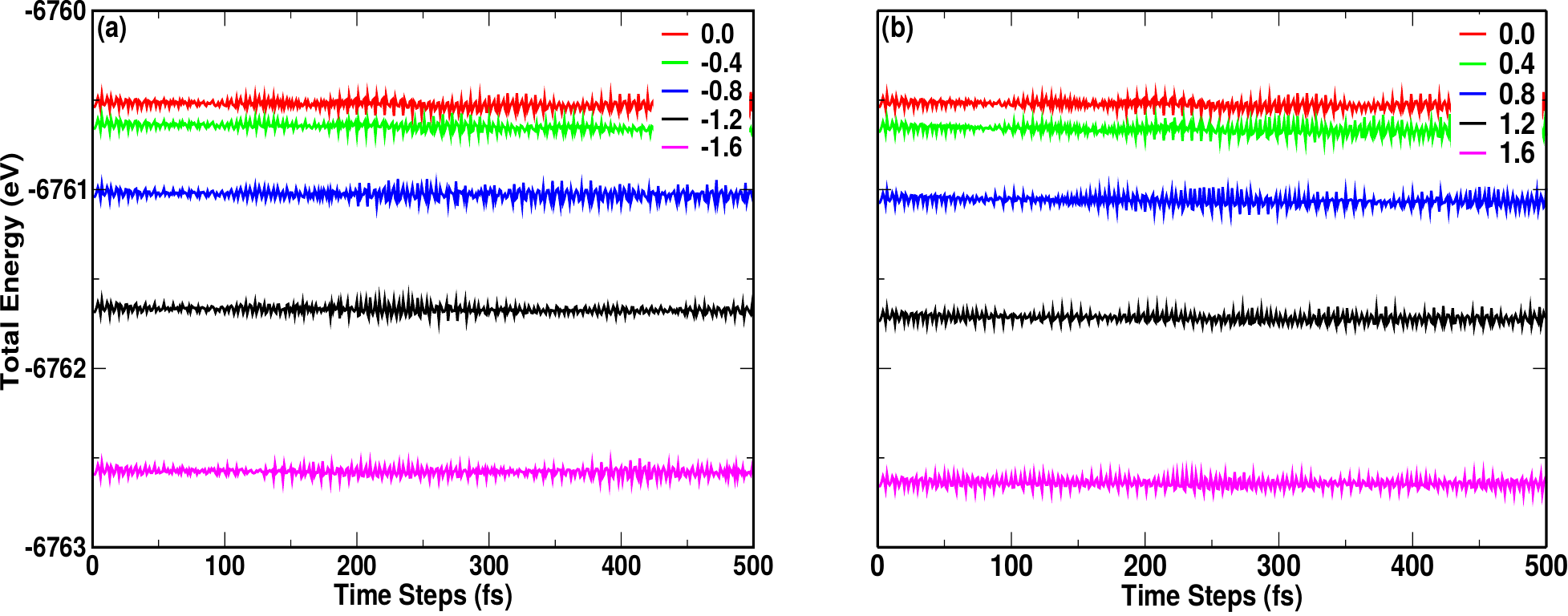}
					\caption{\label{fig:MD} Molecular Dynamics plots for the hydrogen adsorbed h-BN bilayer system under the influence of (a) negative[$E_F$(-ve)] and (b) positive[$E_F$(+ve)] external electric field at temperature(T) = 300 K.}
				\end{figure*}
We have performed an ab initio molecular dynamics simulation using Verlet algorithm in SIESTA to check the thermodynamic stability of H$_2$ molecules adsorbed h-BN bilayer system under external electric field in negative and positive directions. The total energy vs time steps plot at finite temperature of 300 K has been presented in Fig.\ref{fig:MD}[a,b] under the influence of external electric field in (a) negative and (b) positive directions. We have performed the molecular dynamics(MD) simulation of hydrogen adsorbed h-BN bilayer system for 500 fs with 1fs time step. The change in total energy w.r.t time step is very negligible as seen from Fig.\ref{fig:MD}[a,b] which conforms the stability of the H$_2$ adsorbed h-BN bilayer system. Also, no structural deformation or clustering of the H$_2$ molecules are observed under the application of external electric field of different intensity.

\section{Conclusion}
In this study, DFT calculations have been performed to study the effect of external electric field on the hydrogen storage properties of h-BN bilayer. The thermodynamic stability of the H$_2$ adsorbed h-BN bilayer system was studied using ab initio molecular dynamics using Verlet algorithm in SIESTA. The electronic properties of the hydrogen molecules adsorbed h-BN bilayer system is modified by the external electric field. The energy band gap has been tuned from 4.31 eV(Zero $E_F$) to 0.70(0.67) eV at 1.6(-1.6) V/{\AA} under the application of external electric field. Whereas, the hydrogen adsorbed h-BN bilayer shows a metallic character at $\pm$ 2.0 V/{\AA}. In a external field-free condition, h-BN bilayer has gravimetric density of 6.7 wt\% with 16H$_2$ adsorbed molecules and has an average adsorption energy of 0.308 eV/H$_2$ and corresponding desorption temperature $\sim$ 243 K. Whereas, the application of external electric field in the range [-1.6,1.6]V/{\AA} enhances the overall hydrogen adsorption ability of the h-BN bilayer. Average adsorption energy is ranged between 0.311-0.918 eV/H$_2$ with a desorption temperature ranged between 245-725 K. Our study revealed that, binding strength as well as the desorption temperature of the H$_2$ molecules are increased under the application of external electric field.

\begin{acknowledgement}
\textbf{D. P. Rai} thanks Core Research Grant from Department of Science and Technology SERB (CRG DST-SERB, New Delhi India) via Sanction no.CRG/2018/000009(Ver-1).\\
\end{acknowledgement}


\bibliography{library}

\end{document}